\documentclass{ecai} 

\usepackage{latexsym}
\usepackage{amssymb}
\usepackage{amsmath}
\usepackage{amsthm}
\usepackage{booktabs}
\usepackage{enumitem}
\usepackage{graphicx}
\usepackage{color}

\usepackage{hyperref}       
\usepackage{url} 

\newcommand{\BibTeX}{B\kern-.05em{\sc i\kern-.025em b}\kern-.08em\TeX}

\begin{document}


\begin{frontmatter}


\paperid{123} 


\title{A comparative study of Bitcoin and Ripple cryptocurrencies trading using Deep Reinforcement Learning algorithms}


\author[A,B]{\fnms{Dieu-Donné}~\snm{Fangnon}\footnote{Equal contribution.}}
\author[A,B]{\fnms{Armandine Sorel}~\snm{Kouyim Meli}\footnotemark}
\author[A,B]{\fnms{Verlon Roel}~\snm{Mbingui}\footnotemark} 
\author[A,B]{\fnms{Phanie Dianelle}~\snm{Negho}\footnotemark} 
\author[A,B]{\fnms{Regis Konan Marcel}~\snm{Djaha}\footnotemark} 
\author[C]{\fnms{ Lema }~\snm{Logamou Seknewna}}
\address[A]{African Masters of Machine Intelligence (AMMI)}
\address[B]{African Institute for Mathematical Sciences (AIMS), Senegal}
\address[C]{AIMS RIC}



\begin{abstract}
Artificial intelligence (AI) has demonstrated remarkable success across various applications. In light of this trend, the field of automated trading has developed a keen interest in leveraging AI techniques to forecast the future prices of financial assets. This interest stems from the need to address trading challenges posed by the inherent volatility and dynamic nature of asset prices. However, crafting a flawless strategy becomes a formidable task when dealing with assets characterized by intricate and ever-changing price dynamics. To surmount these formidable challenges, this research employs an innovative rule-based strategy approach to train Deep Reinforcement Learning (DRL). This application is carried out specifically in the context of trading Bitcoin (BTC) and Ripple (XRP). Our proposed approach hinges on the integration of Deep Q-Network, Double Deep Q-Network, Dueling Deep Q-learning networks, alongside the Advantage Actor-Critic algorithms. Each of them aims to yield an optimal policy for our application. To evaluate the effectiveness of our Deep Reinforcement Learning (DRL) approach, we rely on portfolio wealth  and the trade signal as performance metrics. The experimental outcomes highlight that Duelling and Double Deep Q-Network outperformed when using XRP with the increasing of the portfolio wealth. 
All codes are available in this  \href{https://github.com/VerlonRoelMBINGUI/RL_Final_Projects_AMMI2023}{\color{blue}Github link}.\\ 

\end{abstract}

\end{frontmatter}


\section{Introduction}

Cryptocurrency markets are notoriously volatile and complex, making them a difficult but appealing playground for algorithmic trading. In recent years, Deep Reinforcement Learning (DRL) has emerged as a promising approach to mastering the intricacies of cryptocurrency trading. DRL methods harness the combined strengths of neural networks and reinforcement learning to enable agents to learn effective trading strategies directly from market data \cite{liu2018practical}. This comparative analysis explores the dynamic intersection of cryptocurrency trading and innovative DRL approaches. It provides a holistic review of diverse DRL methods and their relevance in cryptocurrency markets. As digital assets ascend in the global financial landscape, traders and investors are increasingly embracing AI-powered DRL trading strategies. The central goal of this study is to assess the performance, robustness, and adaptability of different DRL algorithms in cryptocurrency trading settings. To this end, we scrutinize a spectrum of DRL approaches, including Deep Q-Networks (DQN) \cite{mnih2013playing}, Double Deep Q-Networks \cite{van2016deep}, Duelling Deep Q-Networks \cite{wang2016dueling} , and Advantage Actor-Critic (A2C) \cite{huang2022a2c}. Each method's strengths, weaknesses, and suitability for cryptocurrency trading will be rigorously evaluated. Furthermore, this study addresses practical considerations such as data preprocessing, feature engineering, risk management, and model hyperparameters to deliver a comprehensive assessment. By contrasting and comparing these DRL approaches, we aspire to offer insightful insights into their potential to augment trading strategies in the volatile and lucrative cryptocurrency realm. The insights from this comparative analysis can be a valuable asset for traders, investors, and researchers seeking to leverage the power of DRL for cryptocurrency trading. Additionally, it advances the wider dialogue on the convergence of AI and machine learning in financial markets, shedding light on the evolving algorithmic trading landscape in the digital age. \\
Our work is structured as follows: In Section \ref{sec2}, we provide a literature review. Section \ref{sec3} covers the methodology, where we present the formalization of the Markov Decision Process, RL algorithms. In Section \ref{sec4}, we discuss the experiments, starting with a description of the dataset, followed by the presentation of the environment, and  present the results.


\section{Literature Review}

\label{sec2}

In this section, we review some classical trading strategies and discuss how RL has been applied
to this field.

Algorithmic trading is a systematic methodology characterized by mathematical modeling and automated execution. It encompasses a variety of trading strategies, such as trend-following \cite{mandal2022kg}, mean-reversion \cite{fatah2022hierarchical}, statistical arbitrage \cite{deugirmenciouglu2023forecasting}, and delta-neutral trading strategies \cite{hung2024intelligent}. In this context, our primary focus is on the evaluation of time series momentum strategies presented in the work of \cite{zohren2023learning}, which serves as a benchmark for our models.

The research conducted by \cite{zohren2023learning}has produced an exceptionally robust trading strategy, simply based on utilizing the sign of returns over the preceding year as a signal. Their study demonstrated the profitability of this approach across a span of 25 years, encompassing 58 different liquid financial instruments.

The existing literature on Reinforcement Learning (RL) in trading can be broadly classified into three primary methodologies: critic-only, actor-only, and actor-critic approaches \cite{sun2023reinforcement}. Notably, the critic approach, predominantly implemented using Deep Q-Networks (DQN), has garnered the most attention in this domain \cite{huang2022achieving, giorgireinforcement}. This approach involves the construction of a state-action value function, denoted as Q, which quantifies the quality of a specific action within a given state.

Actor-only approaches directly optimize the objective function without the need to compute the expected outcomes of each action in a given state. This direct policy learning makes actor-only methods versatile and applicable to continuous action spaces. Notably, in the research conducted by \cite{poh2022enhancing,lu2023evaluation}, offline batch gradient ascent techniques are employed to optimize objective functions like profits or the Sharpe ratio. These approaches are advantageous because they offer an end-to-end differentiable optimization process.

It's important to distinguish this from standard RL actor-only approaches, where the focus is on learning a policy distribution. In these cases, the Policy Gradient Theorem \cite{balakrishnan2022model} and Monte Carlo methods \cite{metropolis1949monte}come into play during training. Models are updated iteratively, typically at the end of each episode, in order to study and refine the distribution of the policy. This distinction highlights the various strategies employed in actor-only RL methods, depending on the specific objectives and challenges encountered in trading scenarios.

The actor-critic approach represents the third category of RL methodologies and addresses the challenges posed by real-time policy updates. This approach hinges on a fundamental concept: the simultaneous updating of two distinct models. The "actor" model governs an agent's actions based on the current state, while the "critic" model assesses the quality or goodness of the chosen actions.

However, it's worth noting that within financial applications, the actor-critic approach has received relatively limited attention compared to other methods. There have been fewer studies in this domain, with only a few notable works, such as those by \cite{balakrishnan2022model,liu2018practical}, exploring its potential and applicability. Despite being less studied, the actor-critic approach holds promise for addressing real-time learning challenges in financial contexts.


\section{Methodology}
\label{sec3}
We present several configurations, which include state and action spaces as well as reward functions. In our study, we will employ four reinforcement learning (RL) algorithms:  Deep Q Networks, Double Deep Q Networks , Dueling Deep Q Networks  and Deterministic Deep Q Networks.

\subsection{Markov Decision Process Formalisation}
We can frame the trading problem as a Markov Decision Process (MDP) in which an agent engages with the environment during discrete time intervals. At each time step t, the agent is provided with a representation of the environment referred to as a state St. Given this state, the agent selects an action $A_t$, and as a consequence of this action, a numerical reward $R_{t+1}$ is assigned to the agent at the subsequent time step, placing the agent in a new state $S_t+1$. The interaction between the agent and the environment generates a trajectory $\tau =  [S_0, A_0, R_1, S_1, A_1, R_2, S_2, A_2, R_3, \cdots  ] $. $ A_t $  any given time step t, the objective of Reinforcement Learning (RL) is to maximize the expected return, denoted as $G_t$ at time t \cite{zhang2020deep}, which essentially represents the expected cumulative rewards, often discounted :

\begin{equation}
    G_t = \sum_{k=t+1}^{T} \gamma^{k-t-1}R_{k}
\end{equation}
When the discounting factor, denoted as $\gamma $, is considered, optimizing the expectation $ \mathbb{E}(G)$ is equivalent to optimizing our expected wealth if the utility function in Equation 1 takes a linear form and we use $R_t$ to represent trade returns.

\subsection{RL Algorithms}
In this section, we  present some algorithms used in this work.
\subsubsection*{Deep Q Networks}
Deep Q-learning Networks (DQN),  employ a neural network to approximate the state-action value function, also known as the $Q$  function. This $Q$  function is used to estimate how advantageous it is for the agent to take a specific action in a particular state \cite{mnih2013playing}. Assuming that our $Q$  function is represented by a set of parameters denoted as $\theta$ , our objective is to minimize the mean squared error between the current $ Q$ value and the target $Q$ value. This minimization process leads to the derivation of the optimal state-action value function.

\begin{equation}
\begin{split}
L(\theta) &= \mathbb{E}\left[(Q_{\theta}(S,A) - Q_{\theta'}(S,A))^2\right] \\
Q_{\theta'}(S_t,A_t) &= r + \gamma \max_{A'} Q_{\theta'}(S_{t+1},A_{t+1})
\end{split}
\end{equation}

where $ L(\theta)$ is the objective function. A problem is that the training of a vanilla DQN is not stable and suffers from variability. Many improvements have been made to stabilise the training process.

\subsubsection*{Double Deep-Q Networks}
The DQN algorithm employs a max operator to estimate the Q-target, deliberately opting for the highest value. This approach not only executes the action but also assesses it based on this particular methodology. However, it has been demonstrated that the selection and evaluation of the highest value tend to be overly optimistic, potentially leading to training stagnation. To address this issue, the Double Deep Q-Network (DDQN) \cite{vanhasselt2015deep} introduces a separation between action selection and evaluation. In this revised process, the action is taken based on a network with parameters represented as $\theta$ , while the action is evaluated using a separate network with parameters denoted as $\theta^{\prime}$  that considers the next state. This distinction can be formally expressed as follows \cite{nips-2010}:

\begin{equation}
y_i=r+\gamma Q\left(s^{\prime}, \underset{a^{\prime} \in A}{\operatorname{argmax}} Q\left(s^{\prime}, a, \theta_i\right) ; \theta_i^{\prime} \mid s, a\right) .
\end{equation}

The DQN algorithm already uses a second network (target network) with weights $\theta^{\prime}$, which can be viewed as a natural choice for the DDQN algorithm. In conclusion, the DDQN is an extension of the DQN, with the key feature that it additionally uses the target network to separate the execution and evaluation process of an action.

\subsubsection*{ Dueling Deep Q Networks}
The Agent's underlying algorithm is the core module of the RL setup. To simplify, the RL agent learns the sequence of actions that maximize an objective function instead of minimizing it as in the case of a typical deep learning pipeline \cite{wang2016dueling}. We do it recursively using the Bellman Equation.
$$
Q(s, a ; \theta)=r+\gamma Q\left(s^{\prime}, \operatorname{argmax}_{a^{\prime}} Q\left(s^{\prime}, a^{\prime} ; \theta\right) ; \theta^{\prime}\right)
$$
Bellman Equation for Deep Q Networks
The $Q(s, a; \theta)$ denotes the maximum expected future reward for choosing action $\boldsymbol{a}$ in state $s$. The $ Q$-value is constantly updated through an iterative process. Deep Q Networks comprise neural networks acting as function approximations for Q-Table. Inside a DQN, the neural network takes the state as input and outputs the $ Q$-value for each action. The action with the maximum value is then chosen and communicated back to the environment.
Dueling DQN is an extension of Deep Q Networks which includes the calculation of the Advantage of action over other actions in the final output layer\cite{wang2016dueling}.
\begin{align}
    Q(s, a)=V(s)+\left(A(s, a)-\frac{1}{|\mathcal{A}|} \sum_{a^{\prime}} A(s, a)\right)
\end{align}






\subsubsection*{Advantage Actor-Critic (A2C)}
The A2C is proposed to solve the training problem of PG by updating the policy in real-time. It consists of two models: one is an actor network that outputs the policy and the other is a critic network that measures how good the chosen action is in the given state \cite{huang2022a2c}. We can update the policy network $\pi(A \mid S, \theta)$ by maximising the objective function:
\begin{align}
J(\theta)=\mathbb{E}\left[\log \pi(A \mid S, \theta) A_{a d v}(S, A)\right]
\end{align}
where $A_{a d v}(S, A)$ is the advantage function defined as:
\begin{equation}
    A_{a d v}\left(S_{t}, A_{t}\right)=R_{t}+\gamma V\left(S_{t+1} \mid w\right)-V\left(S_{t} \mid w\right)
\end{equation}
In order to calculate advantages, we use another network, the critic network, with parameters $w$ to model the state value function $V(s \mid w)$, and we can update the critic network using gradient descent to minimize the TD-error:
\begin{equation}  
J(w)=\left(R_{t}+\gamma V\left(S_{t+1} \mid w\right)-V\left(S_{t} \mid w\right)\right)^{2}
\end{equation}
The $\mathrm{A} 2 \mathrm{C}$ is most useful if we are interested in continuous action spaces as we recude the policy variance by using the advantage function and update the policy in real-time. The training of $\mathrm{A} 2 \mathrm{C}$ can be done synchronously or asynchronously (A3C).


\section{Experiments}
\label{sec4}
\subsection{Description of Dataset}

The data used in this work was downloaded from the yahoo finance. The dataset is a financial dataset containing daily stock market data for multiple assets such as equities, ETFs, and indexes. It spans from August 30, 2015 to August 30, 2023, and contains 1257 rows and 7 columns namely: 

\begin{itemize}
\item Date: The date on which the stock market data was recorded.
\item Open: The opening price of the asset on the given date.
\item High: The highest price of the asset on the given date.
\item Low: The lowest price of the asset on the given date.
\item Close: The closing price of the asset on the given date.
\item Adj Close: The adjusted closing price of the asset on the given date. 
\item Volume: The total number of shares of the asset that were traded on the given date.
\end{itemize}

To train the different models, we have divided our datasets into Two parts: 0.9 for the training and 10 for the test.

\subsection{Environment}
We will now proceed to define the three primary attributes of the trading agent: the state space, the action space, and the reward function.

\subsubsection*{State Space}
In the realm of literature, various attributes have been employed to define state spaces. Notably, historical price data of a security is a consistent inclusion, alongside frequent utilization of associated technical indicators \cite{vanhasselt2015deep}. In our research, we adopt a state representation that encompasses historical prices, returns ($r_{t}$) computed across different time horizons, and technical indicators, including Moving Average Convergence Divergence (MACD) \cite{baz2015dissecting} and the Relative Strength Index (RSI) \cite{wilder1978new}. For each specific time step, we aggregate the past 60 observations for each of these features to create a unified state. Here is a list of the features we incorporate:

\begin{itemize}
  \item Normalised close price series,

  \item Returns over the past month, 2-month, 3-month and 1-year periods are used. Following \cite{lim2019enhancing}, we normalise them by daily volatility adjusted to a reasonable time scale. As an example, we normalise annual returns as $r_{t-252, t} /\left(\sigma_{t} \sqrt{252}\right)$ where $\sigma_{t}$ is computed using an exponentially weighted moving standard deviation of $r_{t}$ with a 60-day span,

  \item MACD indicators are proposed in \cite{baz2015dissecting} where:

\end{itemize}

$$
\begin{aligned}
& \operatorname{MACD}_{t}=\frac{q_{t}}{\operatorname{std}\left(q_{t-252: t}\right)} \\
& q_{t}=(m(S)-m(L)) / \operatorname{std}\left(p_{t-63: t}\right)
\end{aligned}
$$

where $\operatorname{std}\left(p_{t-63: t}\right)$ is the 63-day rolling standard deviation of prices $p_{t}$ and $m(S)$ is the exponentially weighted moving average of prices with a time scale $S$,

\begin{itemize}
  \item The RSI is an oscillating indicator moving between 0 and 100. It indicates the oversold (a reading below 20) or overbought (above 80 ) conditions of an asset by measuring the magnitude of recent price changes. We include this indicator with a look back window of 30 days in our state representations.
\end{itemize}

\subsubsection*{Action Space}

The action space defines the spectrum of actions available to our agent based on the state representation. These actions are as follows:
\begin{itemize}
    \item  -1 = Sell the asset,
    \item  0 = Take no action,
    \item 1 = Buy the asset .
\end{itemize}
The agent conveys its intention to either Buy or Sell by selecting a value from the set $\{-1, 0, 1\}$ as defined above. The method used to interpret these action values depends on the specific algorithm employed by the agent, which we will delve into in greater detail in the subsequent section dedicated to the Agent.
\subsubsection*{Reward Function}
\textbf{Running Rewards:} Running Rewards are given by the environment to the agent as long as the state is non-terminal. The environment rewards the agent based on the action it takes. Let's demonstrate this with an example:
\begin{itemize}
    \item Let the future return for the time period $t$ be $r(t)$
    \item $A(t)$ be the agent's action at time $t . A(t)$ can take the values $\{-1,0,1\}$
    \item $S(t)$ be the vector representation of the state for time $t$
\end{itemize}
Then the reward $R(t)$ which will be received by the agent after taking action $A(t)$ on observing $S(t)$ can be computed as:
\begin{equation}
R(t)=r(t)^* A(t)-|(A(t)-A(t-1))| * C
\end{equation}

Where $ r(t)$ is the future return for the asset and $ C $ represents transaction costs which we are assuming will be in the range of $1-5$ basis points per trade. $(1$ basis point $=1 \% $ of $1 \% =0.0001)$.  For example, if $r(t)$ is positive(negative) and the agent chooses $A(t)$ as $1(-1)$, i.e., $r(t)$ and $A(t)$ have the same sign, then the return is positive and the agent is rewarded. Conversely, if $r(t)$ and $A(t)$ have opposite signs, i.e., if the future return is negative and the agent decides to buy, then the reward will be negative and the agent will be punished with a negative reward. This is a gross oversimplification but it is fundamentally how reinforcement learning agent learns. As the name suggests, the environment Reinforces the decisions made by the agent through positive and negative rewards.\\

\textbf{Terminal Rewards:}
These are the rewards given by the environment when the agent completes the task that is it reaches the terminal state. The rewards given depends on how the Agent reached the terminal state.
\begin{itemize}
    \item If the agent uses up $\mathbf{7 0 \%}$ of the capital, then that is not a very favourable situation for us. So we heavily punish the agent by giving a large negative reward.
    \item If the agent reaches the end of the episode with enough capital, then we will present the agent with a multiple of the final portfolio return. If the final portfolio return is positive, the reward is highly positive and we are teaching the agent that it is learning in the profitable direction. Same for negative returns except that here we are punishing the agent with a large negative reward which will tell the agent to change its strategy.
    \item We can also use a higher negative multiplier to make the agent more riskaverse towards a negative return.
\end{itemize}

\subsection{Experimental Results}
\label{sec5}

Let's now assess our agent's effectiveness by providing him with an initial capital of about \$100,000. Plots of the signal generated and the portfolio wealth accumulated in the test conditions are shown below.  The subplot on the left side displays the trading signals (buy/sell) created by the agent for each of the coins used, including XRP-USD and Bitcoin (BTC-USD), while the subplot on the right side represents the value of the portfolio for differents models.

\subsubsection{Deep Q Network (DQN)}



\begin{figure}[ht] 
\centering

\includegraphics[scale=0.29]{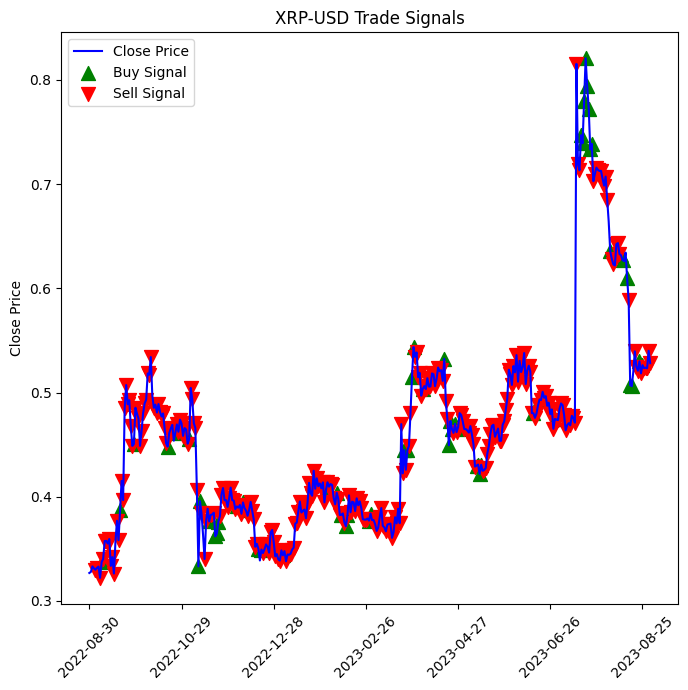}

\includegraphics[scale=0.29]{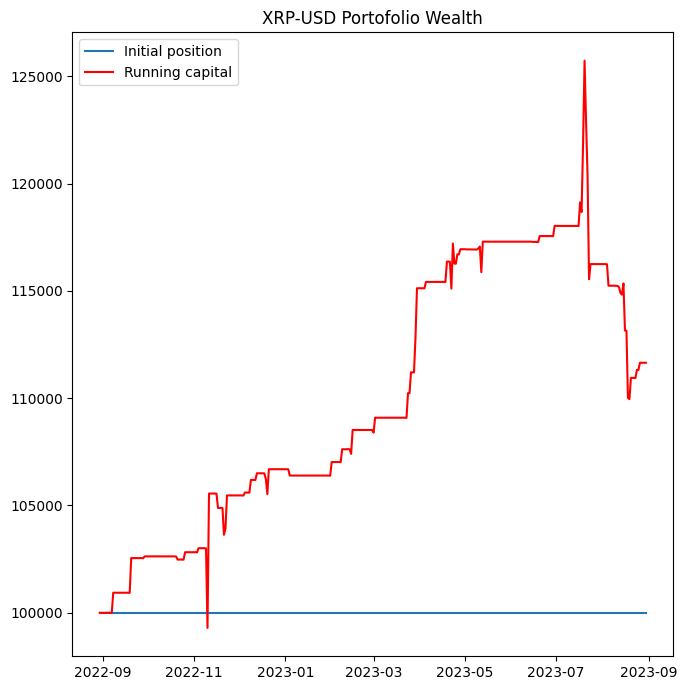}

\caption{XRP USD}
\label{Binance1}
\end{figure}

\begin{figure}[ht] 
\centering

\includegraphics[scale=0.29]{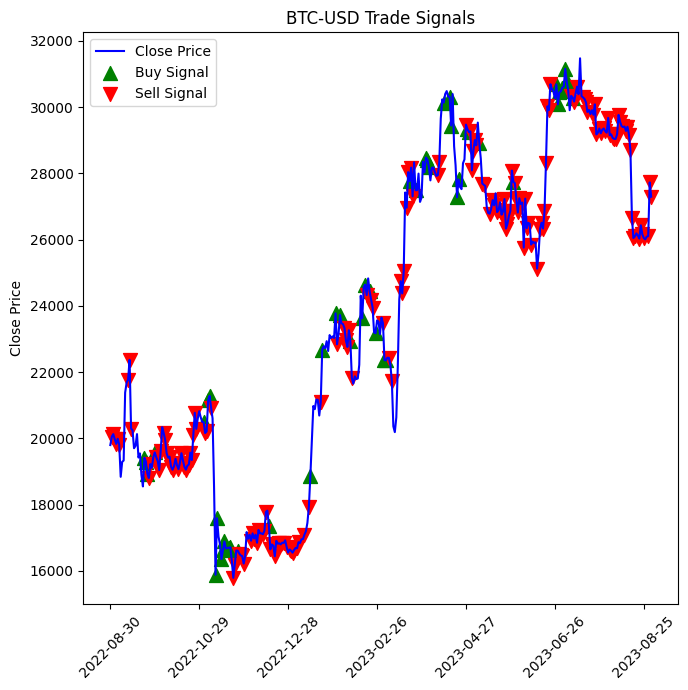}

\includegraphics[scale=0.29]{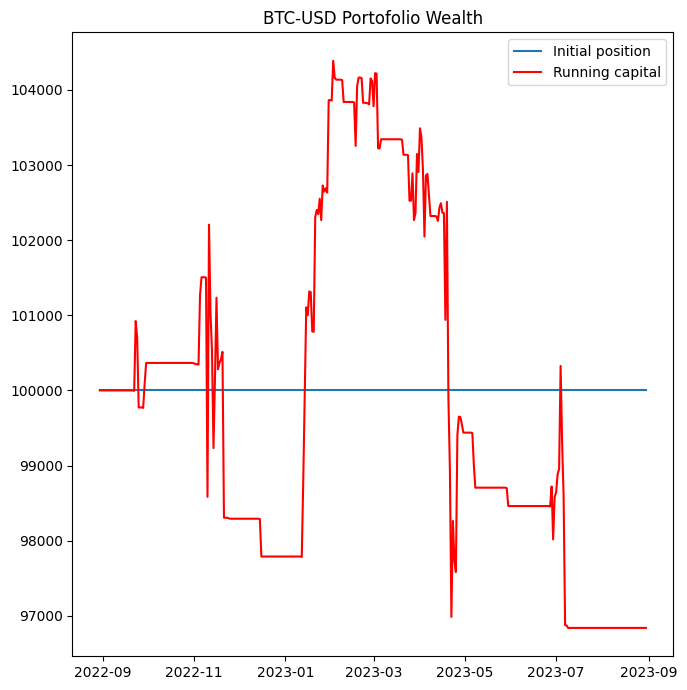}

\caption{Bitcoin USD}
\label{Binance2}
\end{figure}

Here we have trained the agent using the DQN model [\ref{Binance1},\ref{Binance2}]. The results of the above graphs show that, the agent generate more profits with XRP asset than with the Bitcoin cryptocurrency. This conclusion is based on our observation of numerous successful sales and purchases along the signal curve. These transactions have led to substantial gains, as reflected in our portfolio's growth beyond the initial capital investment. 

However, our assessment also reveals fluctuations in the portfolio's value, notably in the vicinity of the initial capital. These variations serve as a testament to the agent's ability to generate both profits and losses during specific time periods. 

From November 2022 to January 2023, we observed significant fluctuations in the portfolio, demonstrating the agent's capacity to navigate through volatile market conditions. Similarly, between May 2023 and the present, we observed a notable decline in the value of Bitcoin (BTC), underscoring the challenges and opportunities presented by the cryptocurrency market in 2023. These results shows that our model performs better with XRP-USD asset than with BTC-USD.

\subsubsection{Double Deep Q Network (DDQN)}

\begin{figure}[ht] 
\centering

\includegraphics[scale=0.29]{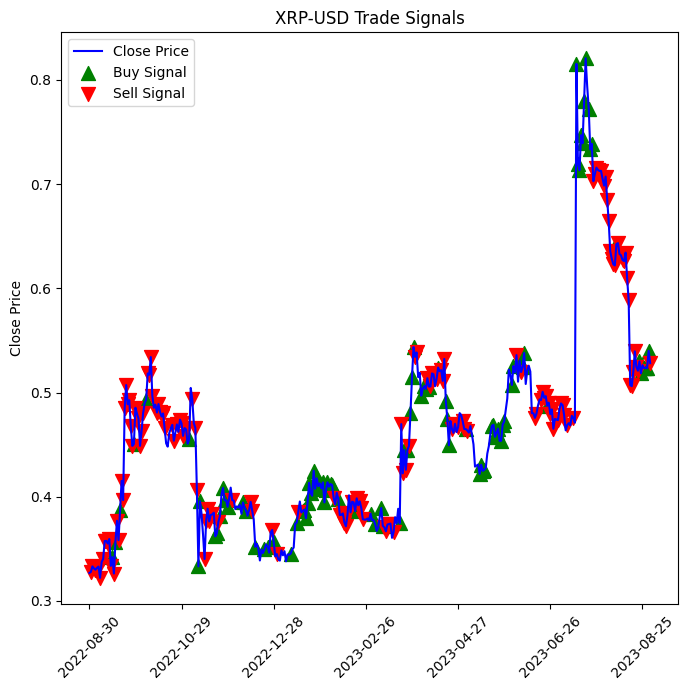}
\label{fig:a}

\includegraphics[scale=0.29]{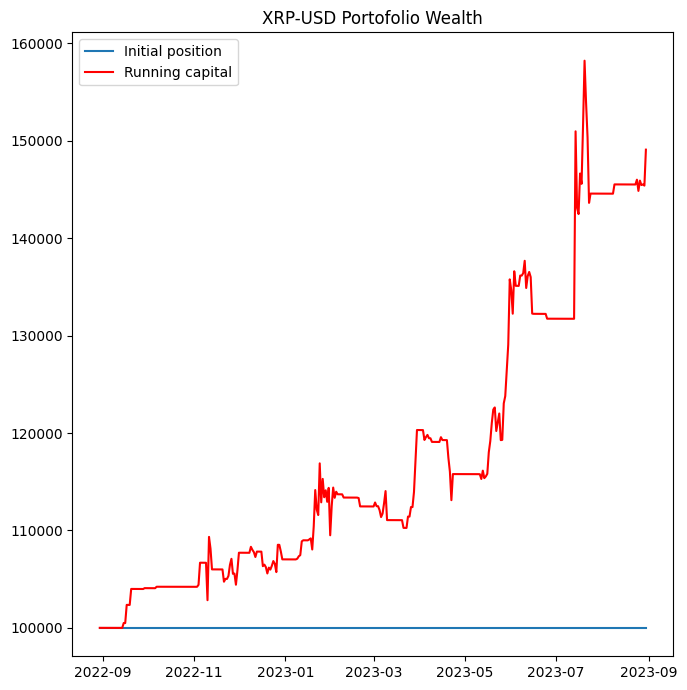}
\label{fig:b}

\caption{XRP USD}
\label{Binance3}

\end{figure}
\vspace*{2cm}
\begin{figure}[ht] 
\centering

\includegraphics[scale=0.29]{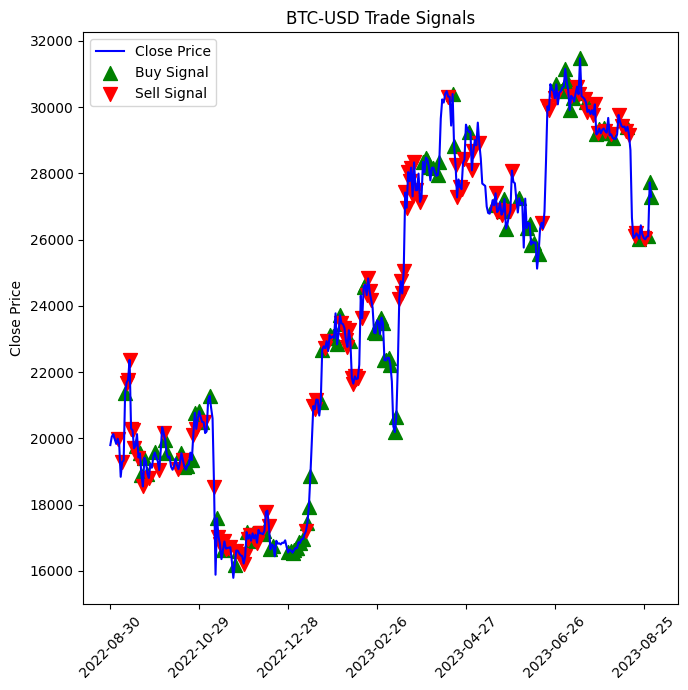}
\label{fig:c}

\includegraphics[scale=0.29]{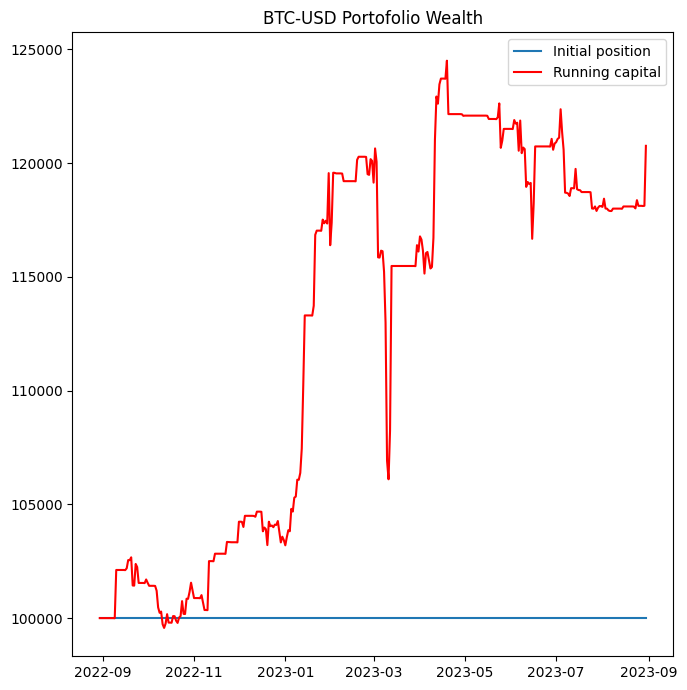}
\label{fig:d}

\caption{Bitcoin USD}
\label{Binance4}
\end{figure}

\clearpage
The provided graphs above [\ref{Binance3},\ref{Binance4}] illustrate the performance of the agent under Double DQN model using Ripple (XRP) and BTC as assets. Firstly, the Ripple portfolio exhibits remarkable growth, 25\% of the initial capital over time, indicating a highly successful investment strategy. Moreover, Ripple's graph indicates consistent gains without any visible losses until 2023. In contrast to the first model, here we see that the agent does generate profits on the initial capital by using Bitcoin with a positive reward. We can therefore conclude that for BTC trading, it is preferable to train an agent with the Double DQN model rather than the single DQN.

\vspace*{1cm}
\subsubsection{Dueling Deep Q Network}

\begin{figure}[ht] 
\centering

\includegraphics[scale=0.29]{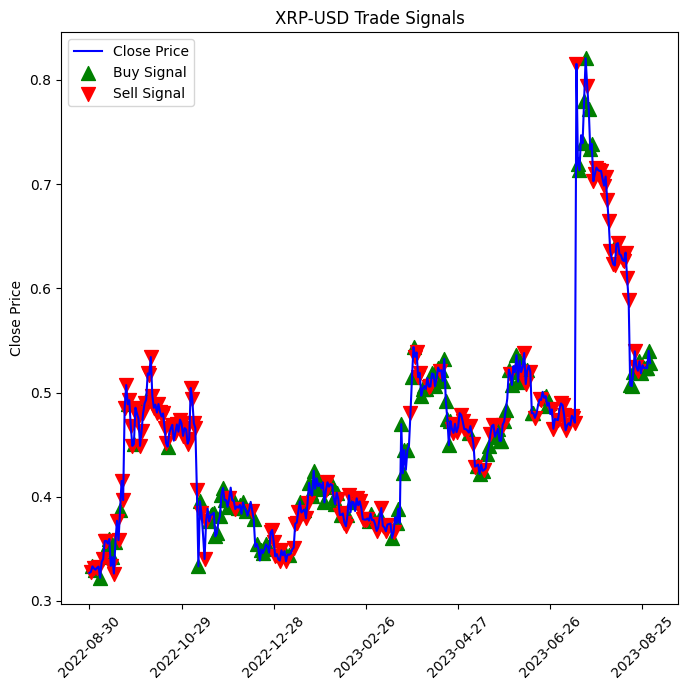}
\label{fig:e}

\includegraphics[scale=0.29]{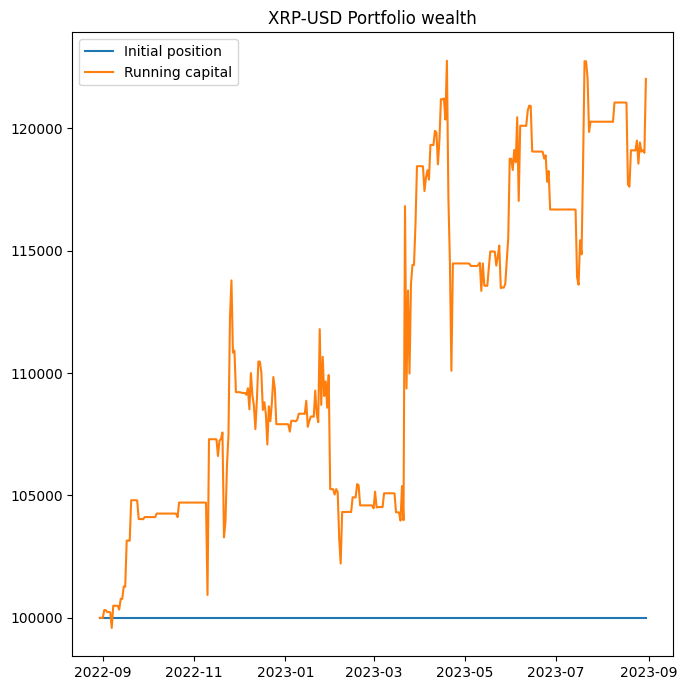}
\label{fig:f}

\caption{Ripple USD}
\label{XRP1}
\end{figure}

\begin{figure}[ht] 
\centering

\includegraphics[scale=0.29]{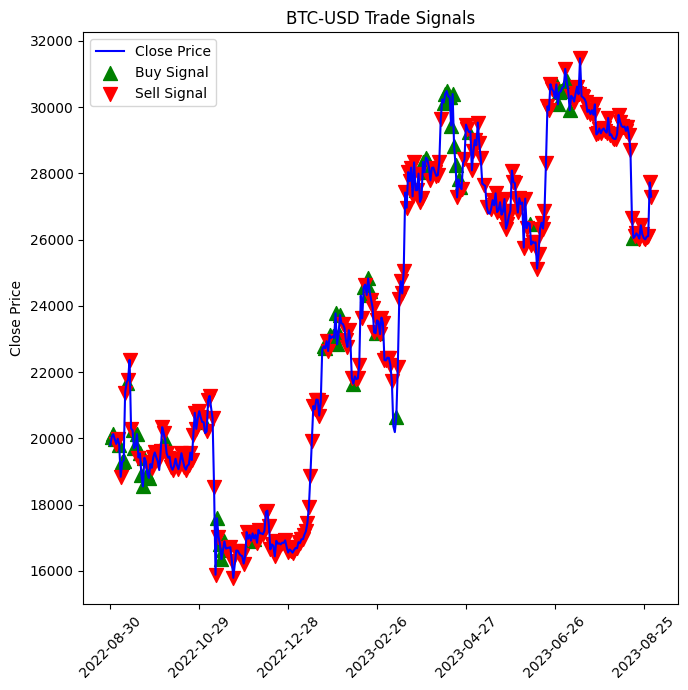}
\label{fig:g}

\includegraphics[scale=0.29]{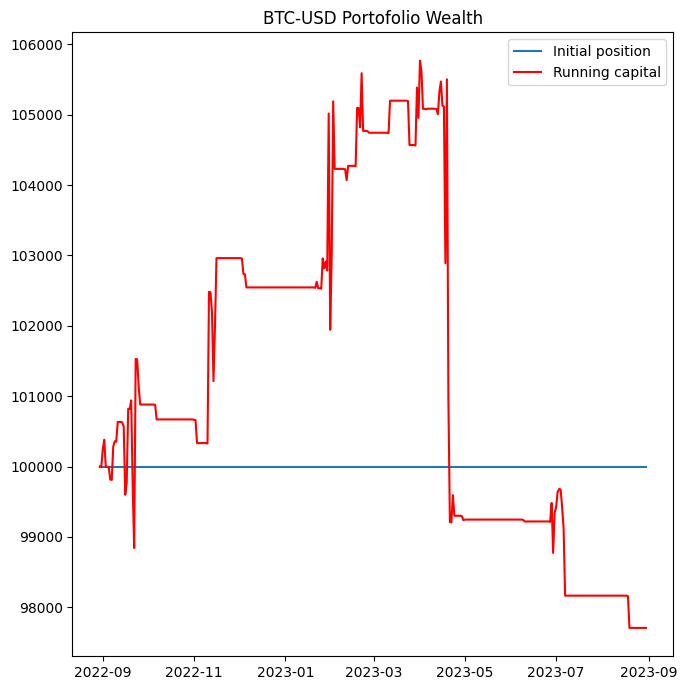}
\label{fig:h}

\caption{Bitcoin USD}
\label{bitcoin1}
\end{figure}

The figures [\ref{XRP1},\ref{bitcoin1}] below show the performance of the agent trained with a new model, the Dueling DQN. With this model, we can clearly see that the agent generates a lot of increasing benefits using XRP. Ripple demonstrates impressive and exponential growth over time, highlighting its potential as a lucrative investment option. Although there were initial losses for several months, the agent's perseverance eventually paid off, showcasing XRP's ability to recover and generate substantial gains. XRP's growth trajectory suggests that it could be one of the best-performing cryptocurrencies, providing investors with opportunities for significant returns. But with Bitcoin, the agent fails to generate profits over the entire period under consideration. At the beginning, the agent generated a large profit, but by May 2023, the initial capital had fallen completely. This downturn serves as a reminder of the volatile nature of cryptocurrencies, where rapid fluctuations can impact investment outcomes. Also, may be due to the volatility of BTC in 2023, the fall in its dollar value on the asset market. So to trade BTC using a Dueling agent, it would be preferable to use data before 2023. But there's no problem with XRP.

\vspace*{1cm}
\subsubsection{Advantage Actor-Critic (A2C)}
\begin{figure}[ht] 
\centering

\includegraphics[scale=0.29]{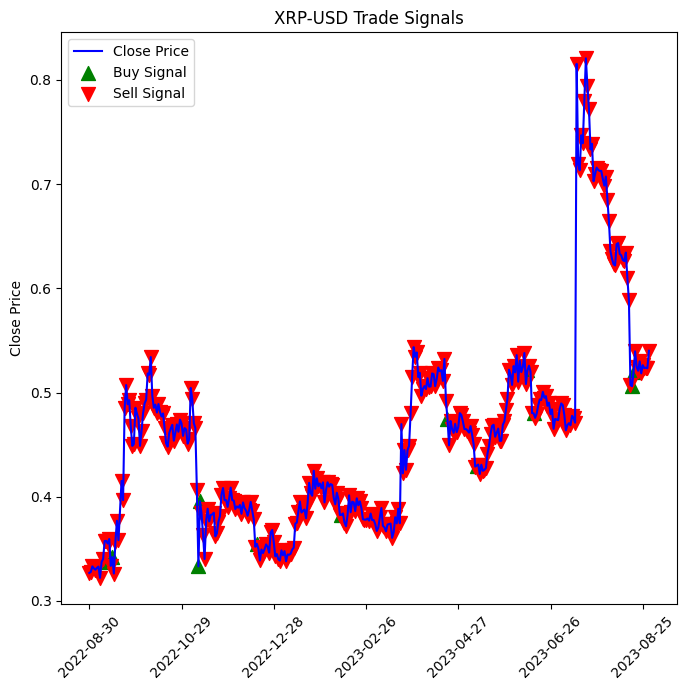}
\label{fig:j}

\includegraphics[scale=0.29]{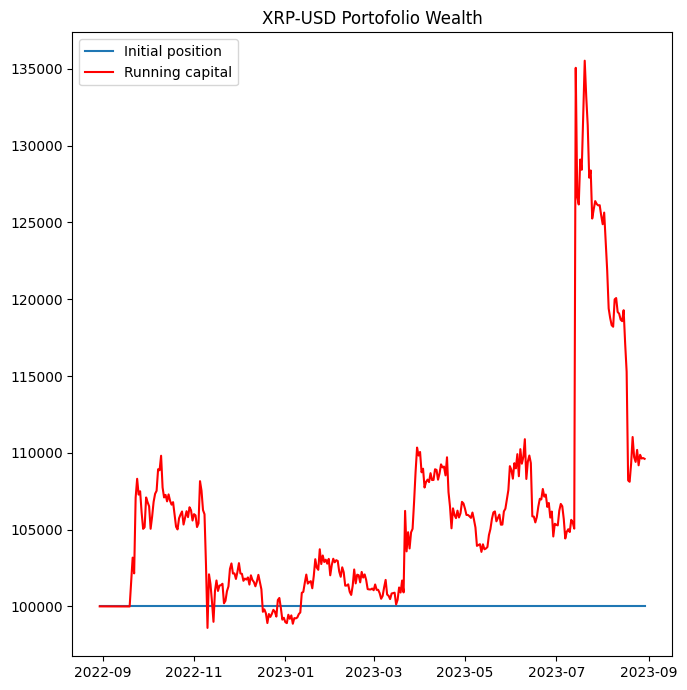}
\label{fig:k}
\caption{XRP-USD}
\label{fig7}
\end{figure}

\newpage

\begin{figure}[ht] 
\centering
\includegraphics[scale=0.29]{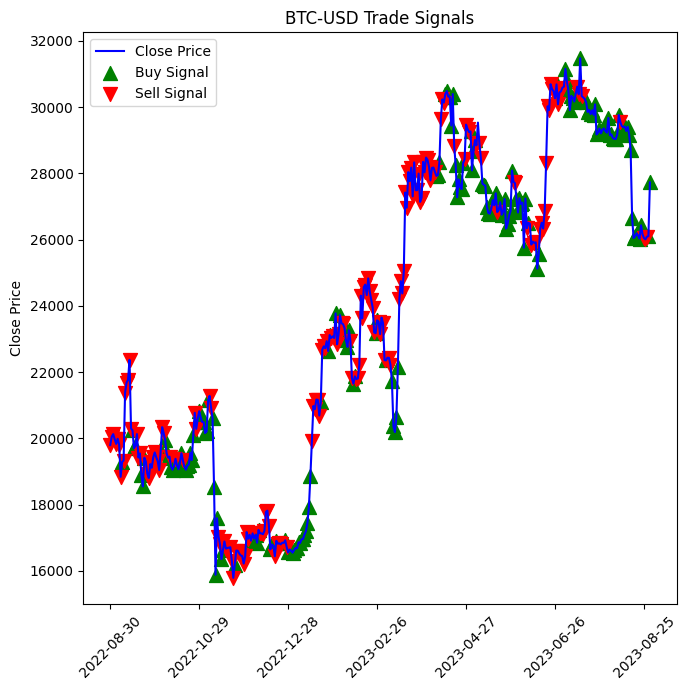}
\label{fig:l}

\includegraphics[scale=0.29]{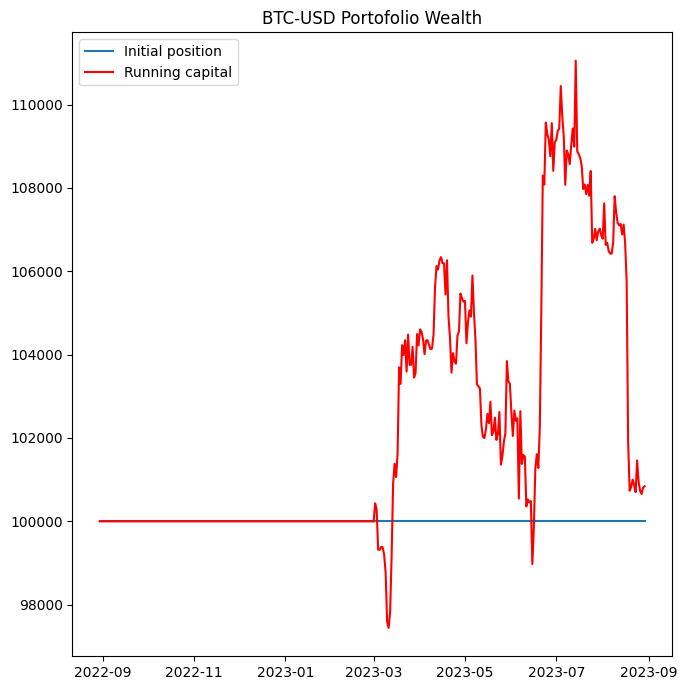}
\label{fig:m}
\caption{BTC-USD}
\label{fig8}
\end{figure}
\vspace*{1cm}
In this case [\ref{fig7},\ref{fig8}], the agent was trained using the A2C model. It can be seen from the XRP graphs above (\ref{fig8}) that the agent sold more than it bought, suggesting that the agent did not generate much profit at the outset. In July 2023, the portfolio grew exponentially before falling sharply again. Compared with other models, A2C is not well suited to XRP trading. In the case of Bitcoin (\ref{fig8}), the agent took no action at the beginning, but capital decreased in March 2023 before growing again and falling again. Of the two models, we can see that the model is better suited to XRP trading than to Bitcoin.


\section{ Conclusion}
In this paper, we delved into the world of cryptocurrency trading, aiming to enhance trading strategies by harnessing the power of deep reinforcement learning (RL). We employed four RL models, namely DQN, Double DQN, Dueling DQN, and A2C, to scrutinize their efficacy in optimizing trading decisions. Our investigation encompassed two prominent cryptocurrencies, XRP and Bitcoin, allowing us to gain insights into the models' performance across different assets. Throughout our experiments, we uncovered valuable insights into the capabilities of these RL agents. Notably, our findings indicated that these models demonstrated superior performance when applied to XRP trading in comparison to Bitcoin. The assessment of their effectiveness was facilitated by the visualization of portfolio wealth plots, where each agent was entrusted with an initial capital and tasked with accumulating profits. Our results underscored the adaptability and potential of deep RL methods in the realm of cryptocurrency trading. While Bitcoin, often regarded as a flagship digital asset, presented challenges and complexities that proved to be formidable for our agents, XRP exhibited a more favorable environment for these models to thrive. It is imperative to acknowledge that the cryptocurrency market is highly dynamic and influenced by multifarious factors, making it a challenging domain for trading algorithms. The differential performance of the RL models across assets highlights the importance of tailoring strategies to the unique characteristics of individual cryptocurrencies. 
\begin{itemize}
    \item The originality of our work in relation to other work on the same subject lies in the fact that we have not only tested the performance of an agent on the financial market for trading by implementing four different models, but we have also extended the range of our data to 2023 (this is a new research study). We were able to show that models such as Double DQN and Dueling DQN perform well for XRP, while for Bitcoin it's Double DQN. This work then forms the basis for future research into the financial market for cryptocurrency trading.
\end{itemize}
To even improve the performance of the agent, we can do some changes as follows:
\begin{itemize}
    \item Use an LSTM Encoder architecture with Attention to extract the features from the time-series data and then pass the feature vector to the agent as input instead of the existing architecture ; Choosing a policy-based method such as the Deep Deterministic Policy Gradient algorithm to specify the amount of asset to buy or sell instead of just going maximally long or short with the investment.
\end{itemize}







\bibliography{paper}

\end{document}